# Dark Energy and Tracker Solution- A Review

Rakhi R., Indulekha K.

School of Pure and Applied Physics

Mahatma Gandhi University, Priyadarshini Hills P.O.

Kottayam, Kerala, India

## Abstract

In this paper, basics and some theoretical models of dark energy are reviewed. Theoretical models include cosmological constant, vacuum fluctuations of quantum fields, scalar field models, chaplygin gas, vector field models and brane world models. Besides this, some alternate models of dark energy are also included. Finally, some elementary ideas on tracker models are also discussed.

**Contents:**

## 1. Introduction: Dark Energy and Dark matter

In the early 1990's, one thing was fairly certain about the expansion of the Universe. It might have enough energy density to stop its expansion and recollapse, it might have so little energy density that it would never stop expanding, but gravity was certain to slow the expansion as time went on. Granted, the slowing had not been observed, but, theoretically, the Universe had to slow. The Universe is full of matter and the attractive force of gravity pulls all matter together. Then came 1998 and the Hubble Space Telescope (HST) observations of very distant supernovae that showed that, a long time ago, the Universe was actually expanding more slowly than it is today. So the expansion of the Universe has not been slowing due to gravity, as everyone thought, it has been accelerating. No one expected this; no one knew how to explain it. But something was causing it. Eventually theorists came up with three sorts of explanations. Maybe it was a result of a long-discarded version of Einstein's theory of gravity, one that contained what was called a "cosmological constant." Maybe there was some strange kind of energy-fluid that filled space. Maybe there is something wrong with Einstein's theory of gravity and a new theory could include some kind of field that creates this cosmic acceleration. Theorists still don't know what the correct explanation is, but they have given the solution a name. It is called dark energy.

**Dark Matter:-**"*a cold, non-relativistic material most likely in the form of exotic particles that interact very weakly with atoms and light*".

In astronomy and cosmology, **dark matter** is hypothetical matter that does not interact with the electromagnetic force, but whose presence can be inferred from gravitational effects on visible matter. According to present observations of structures larger than galaxies, as well as Big Bang cosmology, dark matter and dark energy account for the vast majority of the mass in the observable universe.

The observed phenomena which imply the presence of dark matter include the rotational speeds of galaxies, orbital velocities of galaxies in clusters, gravitational lensing of background objects by galaxy clusters such as the Bullet cluster, and the temperature distribution of hot gas in galaxies and clusters of galaxies. Dark matter also plays a central role in structure formation and galaxy evolution, and has measurable effects on the anisotropy of the cosmic microwave background. All these lines of evidence suggest that galaxies, clusters of galaxies, and the universe as a whole contain far more matter than that which interacts with electromagnetic radiation: the remainder is called the "dark matter component."

The dark matter component has much more mass than the "visible" component of the universe. At present, the density of ordinary baryons and radiation in the universe is estimated to be equivalent to about one hydrogen atom per cubic meter of space. Only about 4% of the total energy density in the universe (as inferred from gravitational effects) can be seen directly. About 22% is thought to be composed of dark matter. The remaining 74% is thought to consist of dark energy, an even stranger component, distributed diffusely in space. Some hard-to-detect baryonic matter is believed to make a contribution to dark matter but would constitute only a small portion. Determining the nature of this missing mass is one of the most important problems in modern cosmology and particle physics.

**2. What is Dark Energy?**

Dark energy is a repulsive force that opposes the self-attraction of matter and causes the expansion of the universe to accelerate. In physical cosmology, dark energy is a hypothetical form of energy that permeates all of space and tends to increase the rate of expansion of the universe. Strangely, dark energy causes expansion because it has strong negative pressure.

A substance has positive pressure when it pushes outward on its surroundings. This is the usual situation for fluids. Negative pressure, or tension, exists when the substance instead pulls on its surroundings. A common example of negative pressure occurs when a solid is stretched to support a hanging weight.

According to the FLRW metric, which is an application of General Relativity to cosmology, the pressure within a substance contributes to its gravitational attraction for other things just as its mass density does. Negative pressure causes a gravitational repulsion.

The gravitational repulsive effect of dark energy's negative pressure is greater than the gravitational attraction caused by the energy itself. At the cosmological scale, it also overwhelms all other forms of gravitational attraction, resulting in the accelerating expansion of the universe.

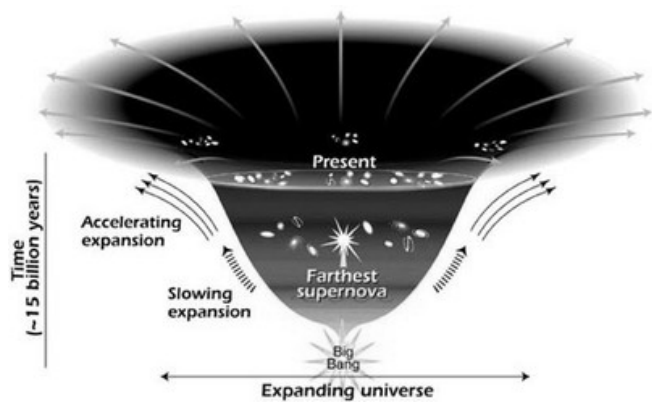

Fig.1 _Universe Dark Energy-1 Expanding Universe_. _This diagram shows changes in the rate of expansion since the Universe's birth 14 billion years ago. The more shallow the curve, the faster the rate of expansion. The curve changes noticeably about 7.5 billion years ago, when objects in the Universe began flying apart at a faster rate._ [Credit: NASA/STSci/Ann Field]

One explanation for dark energy is that it is a property of space. Albert Einstein was the first person to realize that empty space is not anything. Space has amazing properties, many of which are just beginning to be understood. The first property that Einstein discovered is that it is possible for more space to come into existence. Then one version of Einstein's gravity theory, the version that contains a cosmological constant, makes a second prediction: "empty space" can possess its own energy. Because this energy is a property of space itself, it would not be diluted as space expands. As more space comes into existence, more of this energy-of space would appear. As a result, this form of energy would cause the Universe to expand faster and faster. Unfortunately, no one understands why the cosmological constant should even be there, much less why it would have exactly the right value to cause the observed acceleration of the Universe. Another explanation for how space acquires energy comes from the quantum theory of matter. In this theory, "empty space" is actually full of temporary ("virtual") particles that continually form and then disappear. But when physicists tried to calculate how much energy this would give empty space, the answer came out wrong - wrong by a lot. The number came out $10^{120}$ times too big. It's hard to get an answer that bad. Another explanation for dark energy is that it is a new kind of dynamical energy fluid or field, something that fills all of space but something whose effect on the expansion of the Universe is the opposite of that of matter and normal energy.

A last possibility is that Einstein's theory of gravity is not correct. That would not only affect the expansion of the Universe, but it would also affect the way that normal matter in galaxies and clusters of galaxies behaved. This fact would provide a way to decide if the solution to the dark energy problem is a new gravity theory or not: we could observe how galaxies come together in clusters. But if it does turn out that a new theory of gravity is needed, what kind of theory would it be? How could it correctly describe the motion of the bodies in the Solar System, as Einstein's theory is known to do, and still give us the different prediction for the Universe that we need? There are candidate theories, but none are compelling. So the mystery continues.

**3. Evidence for Dark energy**

**i. Supernovae**

In 1998, published observations of Type Ia supernovae by the High-z Supernova Search Team followed in 1999 by the Supernova Cosmology Project suggested that the expansion of the universe is accelerating. Since then, these observations have been corroborated by several independent sources. Measurements of the cosmic microwave background, gravitational lensing, and the large scale structure of the cosmos as well as improved measurements of supernovae have been consistent with the Lambda-CDM model.

Supernovae are useful for cosmology because they are excellent standard candles across cosmological distances. They allow the expansion history of the Universe to be measured by looking at the relationship between the distance to an object and its redshift, which gives how fast it is receding from us. The relationship is roughly linear, according to Hubble's law. It is relatively easy to measure redshift, but finding the distance to an object is more difficult. Usually, astronomers use standard candles: objects for which the intrinsic brightness, the absolute magnitude, is known. This allows the object's distance to be measured from its actually observed brightness, or apparent magnitude. Type Ia supernovae are the best-known standard candles across cosmological distances because of their extreme, and extremely consistent, brightness.

**ii. Cosmic Microwave background (CMB)**

The existence of dark energy, in whatever form, is needed to reconcile the measured geometry of space with the total amount of matter in the universe.

Measurements of cosmic microwave background (CMB) anisotropies, most recently by the WMAP satellite, indicate that the universe is very close to flat. For the shape of the universe to be flat, the mass/energy density of the universe must be equal to a certain critical density. The total amount of matter in the universe (including baryons and dark matter), as measured by the CMB, accounts for only about 30% of the critical density. This implies the existence of an additional form of energy to account for the remaining 70%. The most recent WMAP observations are consistent with a universe made up of 74% dark energy, 22% dark matter, and 4% ordinary matter.

**iii. Large scale structure**

The theory of large scale structure, which governs the formation of structure in the universe (stars, quasars, galaxies and galaxy clusters), also suggests that the density of baryonic matter in the universe is only 30% of the critical density.

**iv. Late time Integrated Sachs-Wolfe Effect (ISW)**

Accelerated cosmic expansion causes gravitational potential wells and hills to flatten as photons pass through them, producing cold spots and hot spots on the CMB aligned with vast super voids and super clusters. This so-called late-time Integrated Sachs-Wolfe effect (ISW) is a direct signal of dark energy in a flat universe.

**4. Candidates of Dark energy**

**i. Cosmological Constant**

The simplest candidate for dark energy is provided by cosmological constant. The cosmological constant corresponds to a fluid with a constant equation of state $w = -1$. But there are certain theoretical issues associate with it: a) the smallest numerical value of lambda leads to fine tuning problem [Fine tuning refers to circumstances when the parameters of the model must be adjusted very precisely in order to agree with observations. Theories requiring fine tuning are regarded as problematic in the absence of a known mechanism to explain why the parameters of the model happen to have precisely the needed values]. b) it leads to coincidence problem[Throughout the history of the Universe, the scalar field density and matter field densities decrease at different rates, so it appears that the conditions in the early universe must be set very carefully in order for the energy densities to be comparable today. This issue of initial conditions is known as "concidence problem"].

*Table* **1: Cosmological Constant Vs. Ordinary Matter**

| **Perfect fluid** of density ρ, Pressure *p*, and 4-velocity *u*. | **Cosmological constant** |
|---|---|
| Energy-momentum tensor $$T_{\mu\nu} = \left(\rho_0 c^2 + p\right) u_\mu u_\nu - p g_{\mu\nu}$$ | Energy-momentum tensor $$T_{\mu\nu} = \frac{c^4}{8\pi G} \Lambda g_{\mu\nu}$$ Energy Density $\rho_\Lambda = \dfrac{c^2}{8\pi G} \Lambda$ |
| Equation of state links density and pressure, e.g., $p = w\rho$, $w = 0$, 1/3 for non-rel. matter, radiation | Classical $\Lambda$ acts as a gas with equation of state $w = -1$; but may also comprise contributions with different equation of state. |
| Adiabatic expansion: non-relativistic matter: $\rho \propto a^{-3}$ radiation: $\rho \propto a^{-4}$ | $\Lambda$ can be a constant during expansion; but more complex contributions evolve differently |
| | All contributions to $\Lambda$ are called "dark energy" |

**ii. A non-zero vacuum energy provided by vacuum fluctuations of**
**quantum fields => quantum field theory**

● Energy associated with space itself (spontaneous creation and destruction of virtual particles; evidence: Casimir effect). Comparison with classical pdV work yields an equation of state with w = - 1.

● Quantum fields can be viewed as a set of harmonic oscillators in momentum space. In their ground state (n=0), these oscillators have a non-zero energy $E_n = \hbar\omega(\vec{k})\left(n + \tfrac{1}{2}\right)$, the total vacuum energy is then given by the sum over all oscillators. The resulting can be transformed into a density of $\rho V \sim \hbar k_{max}^4$, $k_{max}$ is the maximum wave vector of the field, taken to be the energy scale at which QFT fails.

● The Λ problem: take the inverse Planck scale (~$10^{19}$ GeV) for $k_{max}$ => ρV ~ $10^{92}$ gcm$^{-3}$ which is larger than the cosmologically acceptable value by a mere 120 orders of magnitude.

### iii. Scalar Field Models

The cosmological constant corresponds to a fluid with a constant equation of state w = -1. Now, the observations which constrain the value of w today to be close to that of the cosmological constant, these observations actually say relatively little about the time evolution of w, and so we can broaden our horizons and consider a situation in which the equation of state of dark energy changes with time, such as in inflationary cosmology. Scalar fields naturally arise in particle physics including string theory and these can act as candidates for dark energy. So far a wide variety of scalar-field dark energy models have been proposed. These include quintessence, phantoms, K-essence, tachyon, ghost condensates and dilatonic dark energy amongst many. [1][2]

**Quintessence: -** In physics, *quintessence* is a hypothetical form of dark energy postulated as an explanation of observations of an accelerating universe.

Quintessence is a scalar field which has an equation of state (relating its pressure $p_\phi$ and density $\rho_\phi$) of $p_\phi = w_\phi \rho_\phi$, where $w_\phi$ is equal to the equation of state of the energy component dominating the universe (i.e. equal to 1/3 during radiation domination and 0 during matter domination) until $w_\phi$ undergoes a transition to less than -1/3 which initiates the accelerated expansion of the universe. Quintessence is dynamic, and generally has a density and equation of state $\left(w_\phi > -1\right)$ that varies through time and space. By contrast, a cosmological constant is static, with a fixed energy density and $w = -1$.

In quintessence models of dark energy, the observed acceleration of the scale factor is caused by the potential energy of a dynamical field, referred to as quintessence field. Although the *cosmic coincidence* issue remains unresolved, the fine tuning problem facing dark energy/quintessence models with a constant equation of state can be significantly alleviated if we assume that the equation of state is time dependent. Quintessence differs from the cosmological constant in that it can vary in space and time. In order for it not to clump and form structure like matter, the field must be very light so that it has a large Compton wavelength.

The Quintessence field must couple to ordinary matter, which even if suppressed by the Planck scale will lead to long range forces and time dependence of the constants of nature. There are tight constraints on such forces and variations and any successful model must satisfy them.

*Table* 2: **Quintessence (Summary)**

| Action | $S = \int d^4x \sqrt{-g} \left[ -\frac{1}{2}(\nabla\phi)^2 - V(\phi) \right]$ |
|---|---|
| Equation of motion | $\ddot{\phi} + 3H\dot{\phi} + \frac{dV}{d\phi} = 0$ |
| Energy-Momentum tensor | $T_{\mu\nu} = \partial_\mu\phi\partial_\nu\phi - g_{\mu\nu}\left[\frac{1}{2}g^{\alpha\beta}\partial_\alpha\phi\partial_\beta\phi + V(\phi)\right]$ |
| Energy Density | $\rho = -T_0^0 = \frac{1}{2}\dot{\phi}^2 + V(\phi)$ |
| Pressure | $p = T_i^i = \frac{1}{2}\dot{\phi}^2 - V(\phi)$ |
| Equation-of-State Parameter | $w_\phi = \frac{p}{\rho} = \frac{\dot{\phi}^2 - 2V(\phi)}{\dot{\phi}^2 + 2V(\phi)}$ |
| Hubble's Constant (H) | $H = \sqrt{\frac{8\pi G}{3}\left[\frac{1}{2}\dot{\phi}^2 + V(\phi)\right]}$ |
| Acceleration | $\frac{\ddot{a}}{a} = -\frac{8\pi G}{3}\left[\dot{\phi}^2 - V(\phi)\right]$ |
| Condition for Acceleration | $\dot{\phi}^2 < V(\phi)$, which means that one requires a flat potential to give rise to an accelerated expansion |

**K-essence: -** Quintessence relies on the potential energy of scalar fields to lead to the late time acceleration of the universe. It is possible to have a situation where the accelerated expansion arises out of modifications to the kinetic energy of the scalar fields. Originally kinetic energy driven inflation, called K-inflation, was proposed (by Armendariz- Picon et al) to explain early universe inflation at high energies. The analysis was extended to a more general Lagrangian (by Armendariz-Picon et al) and this scenario was called "K-essence".In general, K-essence can be defined *as any scalar field with a non-canonical kinetic energy.*

*Table* **3: K-essence (Summary)**

| Action | $S = \int d^4x\sqrt{-g}\left[\frac{1}{2}R + p(\phi, X)\right]$, where $X \equiv -\left(\frac{1}{2}\right)(\nabla\phi)^2$ and Lag. Density $p(\phi, X) = f(\phi)\hat{p}(X) = K(\phi)X + L(\phi)X^2$ corresponds to a pressure density |
|---|---|
| **Field re-definition** | $\phi_{new} = \int^{\phi_{old}} d\phi \sqrt{\frac{L}{|K|}}$ |
| **Pressure Density (according to the new definition of field)** | $p(\phi, X) = f(\phi)(-X + X^2)$ where $\phi \equiv \phi_{new}$, $X \equiv X_{new} = \left(\frac{L}{|K|}\right)X_{old}$ and $f(\phi) = \frac{K^2(\phi_{old})}{L(\phi_{old})}$ |
| **Energy Density** | $\rho = 2X\frac{\partial p}{\partial X} - p = f(\phi)(-X + 3X^2)$ |
| **Equation-of-State Parameter** | $w = \frac{p}{\rho} = \frac{1-X}{1-3X}$ |
| **Condition for Acceleration** | $w_\phi < -\frac{1}{3}$, which translates into the condition $X < \frac{2}{3}$ |

Equation-of state parameter shows that the kinetic term X plays a crucial role in determining the equation of state of φ. As long as X belongs in the range $1/2 < X < 2/3$, the field φ behaves as dark energy for $0 \leq \alpha \leq 2$ where $\alpha = \frac{2(1+w_\phi)}{1+w_m}$, $w_m$ being the equation-of-state of the background fluid, during matter/radiation dominant era.

**Tachyon field: -** Recently it has been suggested that rolling tachyon condensates, in a class of string theories, may have interesting cosmological consequences.

A rolling tachyon has an interesting equation of state whose parameter smoothly interpolates between −1 and 0. This has led to a flurry of attempts being made to construct viable cosmological models using the tachyon as a suitable candidate for the inflaton at high energy. However tachyon inflation in open string models is typically plagued by several difficulties associated with density perturbations and reheating. Meanwhile the tachyon can also act as a source of dark energy depending upon the form of the tachyon potential.

*Table* **4: Tachyon Field (Summary)**

| | |
|---|---|
| **Action** | $S = -\int d^4x V(\phi)\sqrt{-\det(g_{ab} + \partial_a\phi\partial_b\phi)}$ |
| **Tachyon Potential $V(\phi)$** [from open string theory] | $V(\phi) = \dfrac{V_0}{\cosh(\phi/\phi_0)}$ where $\phi_0 = \sqrt{2}$ for non-BPS D-brane in super string and $\phi_0 = 2$ for the bosonic string |
| **Tachyon potential giving the power-law expansion, $a \propto t^p$** | $V(\phi) = \dfrac{2p}{4\pi G}\left(1 - \dfrac{2}{3p}\right)^{1/2} \phi^{-2}$ [Tachyon potentials which are not steep compared to $V(\phi) \propto \phi^{-2}$ lead to an accelerated expansion] |
| **Equation of motion** | $\dfrac{\ddot{\phi}}{1-\dot{\phi}^2} + 3H\dot{\phi} + \dfrac{1}{V}\dfrac{dV}{d\phi} = 0$ |
| **Energy-Momentum tensor** | $T_{\mu\nu} = \dfrac{V(\phi)\partial_\mu\phi\partial_\nu\phi}{\sqrt{1+g^{\alpha\beta}\partial_\alpha\phi\partial_\beta\phi}} - g_{\mu\nu}V(\phi)\sqrt{1+g^{\alpha\beta}\partial_\alpha\phi\partial_\beta\phi}$ |
| **Energy Density** | $\rho = -T_0^0 = \dfrac{V(\phi)}{\sqrt{1-\dot{\phi}^2}}$ |
| **Pressure** | $p = T_i^i = -V(\phi)\sqrt{1-\dot{\phi}^2}$ |
| **Equation-of-State Parameter** | $w_\phi = \dfrac{p}{\rho} = \dot{\phi}^2 - 1$ |
| **Hubble's Constant (H)** | $H = \left[\dfrac{8\pi G V(\phi)}{3\sqrt{1-\dot{\phi}^2}}\right]^{1/2}$ |
| **Acceleration** | $\dfrac{\ddot{a}}{a} = \dfrac{8\pi G V(\phi)}{3\sqrt{1-\dot{\phi}^2}}\left[1 - \dfrac{3}{2}\dot{\phi}^2\right]$ |
| **Condition for Acceleration** | $\dot{\phi}^2 < 2/3$ |

Irrespective of the steepness of the tachyon potential, the equation of state varies between 0 and −1, in which case the tachyon energy density behaves as $\rho \propto a^{-m}$ with $0 < m < 3$.

**Phatom (Ghost) field: -** Recent observational data indicates that the equation of state parameter w lies in a narrow strip around w = −1 and is quite consistent with being below this value. The scalar field models discussed in the previous subsections corresponds to an equation of state w ≥ −1. The region where the equation of state is less than −1 is typically referred to as some form of phantom (ghost) dark energy. Meanwhile the simplest explanation for the phantom dark energy is provided by a scalar field with a negative kinetic energy.

*Table* **5: Phantom field (Summary)**

| Action | $S = \int d^4 x \sqrt{-g} \left[ \frac{1}{2} (\nabla \phi)^2 - V(\phi) \right]$ |
|---|---|
| Equation of motion | $\ddot{\phi} + 3H \dot{\phi} - \frac{dV}{d\phi} = 0$ |
| Energy-Momentum tensor | $T_{\mu\nu} = \partial_\mu \phi \partial_\nu \phi + g_{\mu\nu} \left[ \frac{1}{2} g^{\alpha\beta} \partial_\alpha \phi \partial_\beta \phi - V(\phi) \right]$ |
| Energy Density | $\rho = -T_0^0 = -\frac{1}{2} \dot{\phi}^2 + V(\phi)$ |
| Pressure | $p = T_i^i = -\frac{1}{2} \dot{\phi}^2 - V(\phi)$ |
| Equation-of-State Parameter | $w_\phi = \frac{p}{\rho} = \frac{\dot{\phi}^2 + 2V(\phi)}{\dot{\phi}^2 - 2V(\phi)}$ |
| Hubble's Constant (H) | $H = \sqrt{\frac{8\pi G}{3} \left[ -\frac{1}{2} \dot{\phi}^2 + V(\phi) \right]}$ |
| Acceleration | $\frac{\ddot{a}}{a} = \frac{8\pi G}{3} \left[ \dot{\phi}^2 + V(\phi) \right]$ |

The curvature of the universe grows toward infinity within a finite time in the universe dominated by a phantom fluid.

Thus, a universe dominated by Phantom energy culminates in a future curvature singularity ('Big Rip') at which the notion of a classical space-time breaks down. In the case of a phantom scalar field this Big Rip singularity may be avoided if the potential has a maximum, e.g,

$$V(\phi) = V_0 \left[ \cosh\left(\frac{\alpha\phi}{m_{pl}}\right) \right]^{-1}, \text{ where } \alpha \text{ is constant.}$$

Since the energy density of a phantom field is unbounded from below, the vacuum becomes unstable against the production of ghosts and normal (positive energy) fields. Even when ghosts are decoupled from matter fields, they couple to gravitons which mediate vacuum decay processes of the type: vacuum → 2 ghosts + 2γ.

It was shown by Cline et al. that we require an unnatural Lorenz invariance breaking term with cut off of order ∼ MeV to prevent an overproduction of cosmic gamma rays.

Also phantom fields are generally plagued by severe Ultra-Violet (UV) *quantum instabilities.* Hence the fundamental origin of the phantom field still poses an interesting challenge for theoreticians.

**Dilatonic field: -** It is mentioned in the previous section that the phantom field with a negative kinetic term has a problem with quantum instabilities. Dilatonic model solves this problem. (**Dilaton** is a hypothetical particle that appears in Kaluza-Klein theory and string theory. i.e, a dilaton is a particle of a scalar field φ; a scalar field (following the Klein-Gordon equation) that always comes with gravity). This model is also an interesting attempt to explain the origin of dark energy using string theory. In 2008, A Carbo et.al showed that the form of potential for the *Dilaton* suggests that after solving for the cosmological evolution of the model, the thermal energy of the universe could be gradually transformed in energy of the *Dilaton*, which then could play the role of a quintessence field describing dark energy.

**iv. Chaplygin gas**

So far we have discussed a number of scalar-field models of dark energy. There exists another interesting class of dark energy models involving a fluid known as a Chaplygin gas [2]. This fluid also leads to the acceleration of the universe at late times.

Remarkably, the Chaplygin gas appears like pressure-less dust at early times and like a cosmological constant during very late times, thus leading to an accelerated expansion.The Chaplygin gas can be regarded as a special case of a tachyon with a constant potential.

However it was shown that the Chaplygin gas models are under strong observational pressure from CMB anisotropies. This comes from the fact that the Jeans instability of perturbations in Chaplygin gas models behaves similarly to cold dark matter fluctuations in the dust-dominant stage but disappears in the acceleration stage. The combined effect of the suppression of perturbations and the presence of a non-zero Jeans length gives rise to a strong integrated Sachs-Wolfe (ISW) effect, thereby leading to the loss of power in CMB anisotropies.

### *Table* 6: Chaplygin gas (Summary)

| | |
|---|---|
| **Lagrangian density** | $L = -V_0\sqrt{1-\phi_{,\mu}\phi^{,\mu}}$, where $\phi_{,\mu} \equiv \partial\phi/\partial x^\mu$ |
| **Four-velocity** | $u^\mu = \dfrac{\phi^{,\mu}}{\sqrt{\phi_{,\alpha}\phi^{,\alpha}}}$ |
| **Energy-Momentum tensor** | $T_{\mu\nu} = (\rho+p)u_\mu u_\nu - p g_{\mu\nu}$ |
| **Energy Density** | $\rho = \dfrac{V_0}{\sqrt{1-\phi_{,\mu}\phi^{,\mu}}}$ |
| **Pressure** | $p = -V_0\sqrt{1-\phi_{,\mu}\phi^{,\mu}}$ |
| **Equation-of-State** | $p_c = -A/\rho_c$, where $\rho_c = \sqrt{A+B(1+z)^6}$, z is redshift and $A = B\left\{\kappa^2\left(\dfrac{1-\Omega_m}{\Omega_m}\right)^2 - 1\right\}$ or $A = V_0^2$ |
| **Hubble's Parameter [H(z)]** | $H(z) = H_0\left[\Omega(1+z)^3 + \dfrac{\Omega}{\kappa}\sqrt{\dfrac{A}{B}+(1+z)^6}\right]^{1/2}$, where $\kappa = \rho_{0m}/\sqrt{B}$ |

A 'generalized Chaplygin gas' has also been proposed for which $p \propto -1/p^\alpha$. The equation of state in this case is

$$w(a) = -\dfrac{|w_0|}{\left[|w_0| + \dfrac{1-|w_0|}{a^{3(1+\alpha)}}\right]}$$

, which interpolates between w=0 at early times ($a \ll 1$) and $w = -1$ at very late times ($a \gg 1$). $w_0$ is the current equation of state when $a = 1$. (The constant $\alpha$ regulates the transition time in the equation of state). WMAP, supernovae and large scale structure data have all been used to test Chaplygin gas models.

**v. Vector Field Models**

Nowadays scientists consider several new classes of viable vector field alternatives to the inflation and quintessence scalar fields.

In 2008, Tomi Koivisto and David F. Mota presented a paper entitles "Vector field models of inflation and Dark Energy"[3]. In their work, spatial vector fields are shown to be compatible with the cosmological anisotropy bounds if only slightly displaced from the potential minimum while dominant, or if driving an anisotropic expansion with nearly vanishing quadropole today. The Bianchi I model with a spatial field and an isotropic fluid is studied as a dynamical system, and several types of scaling solutions are found. On the other hand, time-like fields are automatically compatible with large-scale isotropy. They show that they can be dynamically important if non-minimal gravity couplings are taken into account. As an example, they reconstruct a vector-Gauss-Bonnet model which generates the concordance model acceleration at late times and supports an inflationary epoch at high curvatures. The evolution of the vortical perturbations in these models is computed.

Jose Beltrán Jime'nez and Antonio L. Maroto, in 2008, explored the possibility that the present stage of accelerated expansion of the universe is due to the presence of a cosmic vector field [4]. They had showed that vector theories allow for the generation of an accelerated phase without the introduction of potential terms or unnatural scales in the Lagrangian. They proposed a particular model with the same number of parameters as $\Lambda$CDM and excellent fits to SNIa data. The model is scaling during radiation era, with natural initial conditions, thus avoiding the cosmic coincidence problem. They concluded that vector theories offer an accurate phenomenological description of dark energy in which fine-tuning problems could be easily avoided.

**vi. Brane World Models**

Here, the idea rests on the notion that space-time is higher-dimensional, and that our observable universe is a (3+1)-dimensional 'brane' which is embedded in a (4+1)-dimensional braneworld models allow the expansion dynamics to be radically different from that predicted by conventional Einstein's gravity in 3+1 dimensions. Some cosmological 'surprises' which spring from Braneworld models include [1]:

- Both early and late time acceleration can be successfully unified within a single scheme (Quintessential Inflation) in which the very same scalar field which drives Inflation at early times becomes Quintessence at late times
- The (effective) equation of state of dark energy in the braneworld scenario can be 'phantom-like' (w<-1) or 'Quintessence-like' (w>-1).These two possibilities are essentially related to the two distinct ways in which the brane can be embedded in the bulk.

- The acceleration of the universe can be a *transient* phenomenon: braneworld models accelerate during the present epoch but return to matter-dominated expansion at late times.
- A class of braneworld models encounter a *Quiescent future singularity*, at which $\dot{a} \to const$, but $\ddot{a} \to -\infty$. The surprising feature of this singularity is that while the Hubble parameter, density and pressure remain finite, the deceleration parameter and all curvature invariants *diverge* as the singularity is approached.
- A spatially flat Braneworld can *mimick* a closed universe and *loiter* at large redshifts.
- A braneworld embedded in a five dimensional space in which the extra (bulk) dimension is time-like can *bounce at early times,* thereby generically avoiding the big bang singularity. Cyclic models of the universe with successive expansion-contraction cycles can be constructed based on such a bouncing braneworld.

**vii. Alternative ideas**

Some theorists think that dark energy and cosmic acceleration are a failure of general relativity on very large scales, larger than super-clusters. It is a tremendous extrapolation to think that our law of gravity, which works so well in the solar system, should work without correction on the scale of the universe. Most attempts at modifying general relativity, however, have turned out to be either equivalent to theories of quintessence, or inconsistent with observations. It is of interest to note that if the equation for gravity were to approach r instead of $r^2$ at large, intergalactic distances, then the acceleration of the expansion of the universe becomes a mathematical artifact, negating the need for the existence of Dark Energy.

Alternative ideas for dark energy have come from string theory, DGP model, the holographic principle, Gravity corrections etc, but have not yet proved as compelling as quintessence and the cosmological constant.

**String curvature corrections: -** It is interesting to investigate the string curvature corrections to Einstein gravity amongst which the Gauss-Bonnet correction enjoys special status. These models, however, suffer from several problems. Most of these models do not include tracker like solution and those which do are heavily constrained by the thermal history of universe. For instance, the Gauss-Bonnet gravity with dynamical dilaton might cause transition from matter scaling regime to late time acceleration allowing to alleviate the fine tuning and coincidence problems.

**DGP Model: -** In DGP model, gravity behaves as four dimensional at small distances but manifests its higher dimensional effects at large distances. The modified Friedmann equations on the brane lead to late time acceleration. The model has serious theoretical problems related to ghost modes superluminal fluctuations. The combined observations on background dynamics and large angle anisotropies reveal that the model performs worse than $\Lambda - CDM$.

**Non-Local Cosmology: -** An interesting proposal on non-locally corrected gravity involving a function of the inverse d'Almbertian of the Ricci scalar, $f(\Box^{-1}R)$. For a generic function $f(\Box^{-1}R) \sim \exp(\alpha \Box^{-1}R)$, the model can lead to de-Sitter solution at late times. The range of stability of the solution is given by $\frac{1}{3} < \alpha < \frac{1}{2}$ corresponding to the effective EoS parameter $w_{eff}$ ranging as $\infty < w_{eff} < -\frac{2}{3}$. For $\frac{1}{3} < \alpha < \frac{1}{2}$ and $\frac{1}{2} < \alpha < \frac{2}{3}$, the underlying system is shown to exhibit phantom and non-phantom behavior respectively; the de Sitter solution corresponds to $\alpha = \frac{1}{2}$. For a wide range of initial conditions, the system mimics dust like behavior before reaching the stable fixed point at late times. The late time phantom phase is achieved without involving negative kinetic energy fields. Unfortunately, the solution becomes unstable in presence of the background radiation/matter.

$f(R)$ **Theories of gravity: -** On purely phenomenological grounds, one could seek a modification of Einstein gravity by replacing the Ricci scalar by f(R). The f(R) gravity theories giving rise to cosmological constant in low curvature regime are plagued with instabilities and on observational grounds they are not distinguished from cosmological constant.

The action of $f(R)$ gravity is given by $S = \int \sqrt{-g} d^4 x \left[ \frac{f(R)}{16\pi G} + L_m \right]$. The functional form of $f(R)$ should satisfy certain requirements for the consistency of the modified theory of gravity. The stability of $f(R)$ theory would be ensured provided that,

- $f'(R) > 0$ – graviton is not ghost,

- $f''(R) > 0$ – scalaron is not tachyon.

The $f(R)$ models which satisfy the stability requirements can broadly be classified into categories: (i) Models in which $f(R)$ diverge for $R \to R_0$ where $R_0$ finite or $f(R)$ is non analytical function of the Ricci scalar. These models either can not be distinguishable from $\Lambda CDM$ or are not viable cosmologically. (ii) Models with $f(R) \to 0$ for $R \to 0$ and reduce to cosmological constant in high curvature regime. These models reduce to $\Lambda CDM$ in high redshift regime and give rise to cosmological constant in regions of high density and differ from the latter otherwise; in principal these models can be distinguished from cosmological constant.

Unfortunately, the $f(R)$ models with chameleon mechanism are plagued with curvature singularity problem which may have important implications for relativistic stars. The model could be remedied with the inclusion of higher curvature corrections. At the onset, it seems that one needs to invoke fine tunings to address the problem. The presence of curvature singularity certainly throws a new challenge to $f(R)$ gravity models.

**5. Tracker Solution for Dark Energy**

A substantial fraction of the energy density of the universe may consist of quintessence in the form of a slowly rolling scalar field. Since the energy density of the scalar field generally decreases more slowly than the matter energy density, it appears that the ratio of the two densities must be set to a special, infinitesimal value in the early universe in order to have the two densities nearly coincide today. Recently, Steinhardt et.al introduced the notion of ***tracker fields*** to avoid this initial conditions problem. The term *"tracker"* is meant to refer to solutions joining a common evolutionary track, as opposed to following closely the background energy density and equation-of-state. The tracker models are similar to inflation in that they funnel a diverse range of initial conditions into a common final state.

Although tracking is a useful tool to promote quintessence as a likely source of the missing energy in the universe, the concept of tracking as given by Steinhardt et al does not ensure the physical viability of quintessence in the observable universe. It simply provides for synchronized scaling of the scalar field with the matter/radiation field in the expanding universe in such a way that at some stage (undefined and unrelated to observations), the scalar field energy starts dominating over matter and may induce acceleration in the Hubble expansion. Since there is no control over the slow roll-down and the growth of the scalar field energy during tracking, the transition to the scalar field dominated phase may take place much later than observed. Moreover, any additional contribution to the energy density of the universe, such as quintessence, is bound to affect the dynamics of expansion and structure formation in the universe. As such, any physically viable scalar field must comply with the cosmological observations related to helium abundance, cosmic microwave background and galaxy formation, which are the pillars of the success of the standard cosmological model. A realistic theory of tracking of scalar fields must, therefore, take into account the astrophysical constraints arising from the cosmological observations.

**i. Why Tracker Solution?**

To overcome the 'fine tuning' or the 'initial value' problem, the notion of tracker fields was introduced.

A key problem with the quintessence proposal is explaining why $\rho_\phi$ and the matter energy density should be comparable today. There are two aspects to this problem. First of all, throughout the history of the universe, the two densities decrease at different rates; so it appears that the conditions in the early universe have to be set very carefully in order for the energy densities to be comparable today. We refer to this issue of initial conditions as the ''coincidence problem''. The very same issue arises with a cosmological constant as well.

A second aspect, which we call the ''fine-tuning problem,'' is that the value of the quintessence energy density (or vacuum energy or curvature) is very tiny compared to typical particle physics scales. The fine-tuning condition is forced by direct measurements; however, the initial conditions or coincidence problem depends on the theoretical candidate for the missing energy.

Recently, Steinhardt et.al introduced a form of quintessence called ''tracker fields'' which avoids the coincidence problem. It permits the quintessence fields with a wide range of initial values of $\rho_\phi$ to roll down along a common evolutionary track with $\rho_m$ and end up in the observable universe with $\rho_\phi$ comparable to $\rho_m$ at the present epoch. Thus, the tracker fields can get around both the coincidence problem and the fine tuning problem without the need for defining a new energy scale for $\Lambda_{eff}$.

An important consequence of the tracker solutions is the prediction of a relation between $w_\phi$ and $\Omega_\phi$ today. Because tracker solutions are insensitive to initial conditions, both $w_\phi$ and $\Omega_\phi$ only depend on $V(\phi)$. Hence, for any given $V(\phi)$, once $\Omega_\phi$ is measured, $w_\phi$ is determined. In general, the closer $\Omega_\phi$ is to unity, the closer $wQ$ is to 21. However, since $\Omega_m \geq 0.2$ today, there is a sufficient gap between $\Omega_\phi$ and unity that $w_\phi$ cannot be so close to -1. We find that $w_\phi \geq -0.8$ for practical models. This $w_\phi$-$\Omega_\phi$ relation makes the tracker field proposal distinguishable from the cosmological constant.

**ii. Tracker Field:-** *a field whose evolution according to its equation-of-motion converges to the same solution—the tracker solution—for a wide range of initial conditions for the field and its time derivative.*

Tracker fields have an equation-of-motion with attractor-like solutions in the sense that a very wide range of initial conditions rapidly converge to a common, cosmic evolutionary track of $\rho_\phi(t)$ and $w_\phi(t)$. Technically, the tracker solution differs from a classical dynamics attractor solution because neither $\Omega_\phi$ nor any other parameters are fixed in time.

The initial value of $\rho_\phi$ can vary by nearly 100 orders of magnitude without altering the cosmic history. The acceptable initial conditions include the natural possibility of equi-partition after inflation—nearly equal energy density in $\phi$ as in the other 100–1000 degrees of freedom (e.g., $\Omega_{\phi i} \approx 10^{-3}$). Furthermore, the resulting cosmology has desirable properties. The equation-of-state $w_\phi$ varies according to the background equation-of-state $w_B$. When the universe is radiation-dominated ($w_B = \frac{1}{3}$), then $w_\phi$ is less than or equal to 1/3 and $\rho_\phi$ decreases less rapidly than the radiation density. When the universe is matter-dominated ($w_B = 0$), then $w_\phi$ is less than zero and $\rho_\phi$ decreases less rapidly than the matter density. Eventually, $\rho_\phi$ surpasses the matter density and becomes the dominant component. At this point, $\phi$ slows to a crawl and $w_\phi \to -1$ as $\Omega_\phi \to 1$ and the universe is driven into an accelerating phase. These properties seem to match current observations well.

### iii. Family of Tracker Solutions

For a potential $V(\phi) = M^4 \tilde{v}(\phi/M)$ (where $\tilde{v}$ is a dimensionless function of $\phi/M$), there is a family of tracker solutions parameterized by $M$. The value of $M$ is determined by the measured value of $\Omega_m$ today (assuming a flat universe).

### iv. Tracker Equation (with quintessence as an eg.)

Many models of quintessence have a *tracker* behavior, which partly solves the cosmological constant problem. In these models, the quintessence field has a density which closely tracks (but is less than) the radiation density until matter-radiation equality, which triggers quintessence to start having characteristics similar to dark energy, eventually dominating the universe. This naturally sets the low scale of the dark energy.

Although the *cosmic coincidence* issue remains unresolved, the fine tuning problem facing dark energy/quintessence models with a constant equation of state can be significantly alleviated if we assume that the equation of state is time dependent. An important class of models having this property are scalar fields which couple minimally to gravity and whose energy momentum tensor is

$$\rho \equiv -T_0^0 = \frac{1}{2}\dot{\phi}^2 + V(\phi), \quad p \equiv T_i^i = \frac{1}{2}\dot{\phi}^2 - V(\phi) \tag{1}$$

The equation-of-motion for the field $\phi$ is $\ddot{\phi} + 3H\dot{\phi} + V' = 0$ and the equation-of-state is

$w_\phi = \dfrac{p}{\rho} = \dfrac{\dot{\phi}^2 - 2V(\phi)}{\dot{\phi}^2 + 2V(\phi)}$ .It is extremely useful to combine these relations into new form for the equation-of-motion as,

$$\pm \frac{V'}{V} = 3\sqrt{\frac{\kappa}{\Omega_\phi}}\sqrt{1+w_\phi}\left[1 + \frac{1}{6}\frac{d\ln x}{d\ln a}\right] \tag{2}$$

, where $x = \dfrac{(1+w_\phi)}{(1-w_\phi)} = \dfrac{1}{2}\dot{\phi}^2 / V$ is the ratio of the kinetic to potential energy density for

$\phi$ and a prime means a derivative with respect to $\phi$. The $\pm$ sign depends on whether $V' > 0$ or $V' > 0$ respectively. The tracking solution (to which general solutions converge) has the property that $w_\phi$ is nearly constant and lies between $w_B$ and -1. For $1+w_\phi = O(1), \dot{\phi}^2 \approx \Omega_\phi H^2$ and the equation-of-motion [Eqn. (2)] dictates that $\dfrac{V'}{V} \approx \dfrac{1}{\sqrt{\Omega_\phi}} \approx \dfrac{H}{\dot{\phi}}$, for a tracking solution; this is referred to as the "**tracker condition.**" [5]

A scalar field rolling down its potential *slowly* generates a time-dependent $\Lambda$-term since $P \simeq -\rho \simeq -V(\phi)$ if $\dot{\phi}^2 \ll V(\phi)$. Potentials which satisfy $\Gamma \equiv V''V / (V')^2 \geq 1$ have the interesting property that scalar fields approach a common evolutionary path from a wide range of initial conditions. In these so-called `tracker' models the scalar field density (and its equation of state) remains close to that of the dominant background matter during most of cosmological evolution. Tracking behavior with $w_\phi < w_B$ occurs for any potential in which $\Gamma \equiv V''V / (V')^2 > 1$ and is nearly constant $\left[\left|d(\Gamma - 1/Hdt)\right| \ll \left|\Gamma - 1\right|\right]$ over the range of plausible initial $\phi$. This case is relevant to tracker models of quintessence since we want $w_\phi < 0$ today. The range of initial conditions extends from $V(\phi)$ equal to the initial background energy density $\rho_B$ down to $V(\phi)$ equal to the background density at matter-radiation equality, a span of over 100 orders of magnitude. The testing for the existence of tracking solutions reduces to a simple condition on $V(\phi)$ without having to solve the equation-of-motion directly. The condition $\Gamma > 1$ is equivalent to the constraint that $\left|V'/V\right|$ be decreasing as $V$ decreases. These conditions encompass an extremely broad range of potentials.

A good example is provided by the exponential potential $V(\phi) = V_0 \exp\left[-(8\pi)^{1/2} \lambda \phi / M_{pl}\right]$ for which

$$\frac{\rho_\phi}{\rho_B + \rho_\phi} = \frac{3(1+w_B)}{\lambda^2} = \text{constant} < 0.2 \tag{3}$$

$\rho_B$ is the background energy density while $w_B$ is the associated equation of state. The lower limit $\rho_\phi / \rho_{total} < 0.2$ arises because of nucleosynthesis constraints which prevent the energy density in quintessence from being large initially (at $t \sim few$ sec.). Since the ratio $\rho_\phi / \rho_{total}$ remains fixed, exponential potentials on their own cannot supply us with a means of generating dark energy/quintessence at the present epoch. However a suitable modification of the exponential achieves this. For instance the class of potentials

$$V(\phi) = V_0 \left[\cosh \lambda \phi - 1\right]^p \tag{4}$$

has the property that $w_\phi \simeq w_B$ at early times whereas $\langle w_\phi \rangle = (p-1)/(p+1)$ at late times. Consequently Eqn. (4) describes *quintessence* for $p \leq 1/2$ and pressure-less `cold' dark matter (CDM) for $p = 1$.

A second example of a tracker-potential is provided by $V(\phi) = V_0/\phi^\alpha$. During tracking the ratio of the energy density of the scalar field (quintessence) to that of radiation/matter gradually increases $\rho_\phi / \rho_B \propto t^{4/2+\alpha}$ while its equation of state remains marginally smaller than the background value $w_\phi = (\alpha w_B - 2)/(\alpha + 2)$. These properties allow the scalar field to eventually dominate the density of the universe, giving rise to a late-time epoch of accelerated expansion. (Current observations place the strong constrains $\alpha \leq 2$.)

Several of the quintessential potentials listed in **Table 7** have been inspired by field theoretic ideas including super-symmetric gauge theories and super-gravity, pseudo-goldstone boson models, etc [1]. However accelerated expansion can also arise in models with: (i) topological defects such as a frustrated network of cosmic strings ($w \simeq -1/3$) and domain walls ($w \simeq -2/3$); (ii) scalar field Lagrangians with non-linear kinetic terms and no potential term (k-essence); (iii) vacuum polarization associated with an ultra-light scalar field; (iv) non-minimally coupled scalar fields; (v) fields that couple to matter; (vi) scalar-tensor theories of gravity; (vii) brane-world models etc.

*Table* 7

| Quintessence Potential | Reference |
|---|---|
| $V_0 \exp(-\lambda\phi)$ | Ratra & Peebles (1988), Wetterich (1988), Ferreira & Joyce (1998) |
| $m^2\phi^2, \lambda\phi^4$ | Frieman et al (1995) |
| $V_0/\phi^\alpha, \alpha > 0$ | Ratra & Peebles (1988) |
| $V_0 \exp(\lambda\phi^2)/\phi^\alpha, \alpha > 0$ | Brax & Martin (1999, 2000) |
| $V_0 (\cosh\lambda\phi - 1)^p$ | Sahni & Wang (2000) |
| $V_0 \sinh^{-\alpha}(\lambda\phi)$ | Sahni & Starobinsky (2000), Ureña-López & Matos (2000) |
| $V_0 \left(e^{\alpha\kappa\phi} + e^{\beta\kappa\phi}\right)$ | Barreiro, Copeland & Nunes (2000) |
| $V_0 (\exp M_p/\phi - 1)$ | Zlatev, Wang & Steinhardt (1999) |
| $V_0 \left[(\phi - B)^\alpha + A\right] e^{-\lambda\phi}$ | Albrecht & Skordis (2000) |

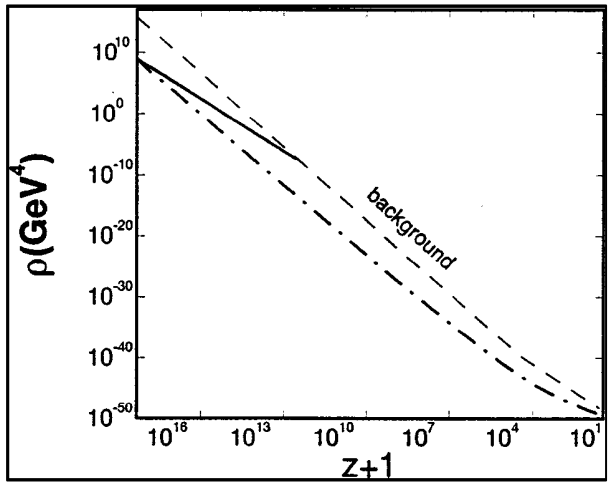

FIG. 2. A plot comparing two tracker solutions for the case of a $V \sim 1/\phi^6$ potential (solid line) and a $V \sim \exp(1/\phi)$ potential (dot dashed line). The dashed line is the background density. The two tracker solutions were chosen to have the same energy density initially.
[Credit: - Steinhardt et.al (1999)]

From the Fig. 2, it is seen that the tracker solution for the generic example [$V \sim \exp(1/\phi)$] reaches the background density much later than for the pure inverse-power law potential. Hence, $\Omega_\phi$ is more likely to dominate late in the history of the universe in the generic case.

Scalar field based quintessence models can be broadly divided into two classes: (i) those for which $\phi/M_{pl} \ll 1$ as $t \to t_0$, (ii) those for which $\phi/M_{pl} \geq 1$ as $t \to t_0$ ($t_0$ is the present time). An important issue concerning the second class of models is whether quantum corrections become important when $\phi/M_{pl} \geq 1$ and their possible effect on the quintessence potential. One can also ask whether a given choice of parameter values is `natural'. Consider for instance the potential $V = M^{4+\alpha}/\phi^\alpha$, current observations indicate $V_0 \simeq 10^{-47} \text{GeV}^4$ and $\alpha \leq 2$, which together suggest $M \lesssim 0.1$ GeV (smaller values of $M$ arise for smaller $\alpha$) it is not clear whether such small parameter values can be motivated by current models of high energy physics.